\documentclass[manuscript]{aastex}
\shorttitle{NSV 11749}
\shortauthors{Bond}

\begin{document}

\title{NSV~11749: Symbiotic Nova, Not a Born-Again Red Giant}

\author{Howard E. Bond\altaffilmark{1,2}
and
Mansi M. Kasliwal\altaffilmark{3,4}}

\altaffiltext{1}{Space Telescope Science Institute;  3700 San Martin Drive,
Baltimore, MD 21218; bond@stsci.edu}

\altaffiltext{2}{Current address: 9615 Labrador Lane, Cockeysville, MD 21030}

\altaffiltext{3}{Observatories of the Carnegie Institution for Science, 813
Santa Barbara St, Pasadena CA 91101; mansi@obs.carnegiescience.edu}

\altaffiltext{4}{Hubble Fellow}

% \clearpage

\begin{abstract}

NSV~11749 is a little-studied variable star, discovered by W. J. Luyten, which
had a long-duration outburst around the year 1903, reaching blue magnitude 12.5
at maximum. Following the outburst, it has apparently been quiescent at about
blue magnitude 17 for the past century. It was recently suggested that NSV~11749
may have been a low- or intermediate-mass star that underwent a final helium
shell flash, making it temporarily a ``born-again'' red giant. If so, it would
be only the fourth known member of this class, along with V605~Aql, FG~Sge, and
V4334~Sgr. However, our newly obtained optical and near-IR spectra of the object
show that it is instead a symbiotic binary, with strong Balmer and He~I--II
emission lines, combined with a cool red-giant companion of spectral type
M1-2~III\null. The 1903 outburst was most likely a symbiotic nova event, of
which less than a dozen are known at present.

\end{abstract}


\clearpage

\section{Introduction: Born-Again Red Giants}

Near the end of its life, a low- or intermediate-mass star leaves the asymptotic
giant branch (AGB) and evolves rapidly across the HR diagram to the top of the
white-dwarf (WD) cooling track. As the star ascends toward the tip of the AGB,
it begins undergoing periodic helium-shell thermal pulses. Depending on the
relative timing of the pulses and the departure from the AGB, the star may
experience a helium final flash after it has already approached or reached
the top of the WD sequence. In this case, the star quickly retraces its
evolution and temporarily becomes a red giant once more, a so-called
``born-again'' red giant. The final-flash scenario was first described theoretically by
Iben et al.\ (1983).

There are three known objects---all of them central stars of extremely faint
planetary nebulae---that appear to have undergone born-again final-flash
events during modern astronomical history: V605~Aquilae and V4334~Sagittarii
(Sakurai's Object), which evolved very rapidly and are generally considered to
represent very late thermal pulses (VLTPs), and FG~Sagittae, which evolved more
slowly and represents a late thermal pulse. There is an extensive literature on
these objects and the underlying theory (e.g., Clayton \& De~Marco 1997; Lawlor
\& MacDonald 2003; Herwig 2005; Sch\"onberner 2008; Bond et al.\ 2013; and
references therein).

\section{NSV 11749}

NSV~11749 is a little-studied variable star that was discovered by Luyten (1937)
in the course of his Bruce Proper Motion Survey, and was originally designated
AN 799.1936 Aquilae.  As recounted by Williams (2005), Luyten found NSV~11749 to
have had blue magnitudes of 13.5 and 17 on plates taken in 1903 and 1934,
respectively. Unfortunately, Luyten never published a finding chart, and the
coordinates that he provided were only approximate. Williams (2005), however,
was able to recover NSV~11749 on plates in the Harvard collection, including the
Bruce plate pair on which Luyten had marked the variable. Williams gave improved
coordinates, as well as a light curve based on 175 Harvard plates obtained
between 1888 and 1988. The star was first detected at blue magnitude 14 in 1899,
and reached a maximum of 12.5 mag in 1903. It remained bright for 4 years,
declined back to 14.5 mag in mid-1911, and then fell below the plate limit apart
from four detections on deep plates at about 17th mag between 1934 and 1949. 
Williams suggested that NSV~11749 might have been a slow nova or an
FU~Orionis-type young stellar object.

However, Miller Bertolami et al.\ (2011) recently made the novel suggestion that
the outburst of NSV~11749 was a VLTP event similar to those undergone by
V605~Aql and V4334~Sgr, on the basis of similarities of the outburst light
curves of the three objects. At the request of these authors, the DASCH team at
Harvard (Grindlay et al.\ 2009) determined an accurate position for NSV~11749
from plates taken during the 1903 outburst. There is a stellar-appearing source
at the DASCH position in various modern sky surveys, including the USNO-NOMAD
catalog (Zacharias et al.\ 2004; J2000 position 19:07:42.36, +00:02:51.0;
$B=16.6$, $R=14.1$), IPHAS (Gonz\'alez-Solares et al.\ 2008; $r'=14.5$,
$i'=13.1$, $H\alpha=13.0$), 2MASS (Cutri et al.\ 2003; $J=10.8$, $H=9.7$,
$K_s=9.4$), and {\it WISE\/} (Wright et al.\ 2010; [3.4] =9.1, [4.6] = 9.1, [12]
= 8.7, [22] = 8.4).  Figure~1 presents a finding chart for this object.

This object was also detected as an H$\alpha$ emission-line star, designated
HBHA -0201-01 ($V\simeq15.1$), in an objective-prism search carried out with
plates obtained with the Hamburg Schmidt telescope in the years 1964--1970
(Kohoutek \& Wehmeyer 1999). The large value of $r'-H\alpha$ in the IPHAS survey
is  consistent with strong H$\alpha$ emission.

There is no known planetary nebula surrounding NSV~11749. The SIMBAD database
identifies NSV~11749 with the bright {\it IRAS\/} source 19050+0001; however,
inspection of the {\it IRAS\/} images shows that the infrared source is actually
a pair of bright stars about $3'$ north of NSV~11749.

\section{Observations}

If the suggestion of Miller Bertolami et al.\ (2011) were correct, NSV~11749
would be only the fourth known born-again event, and would be of particular
interest as a post-FF object that has evolved more than a full century since its
outburst. In order to investigate its nature, we requested spectroscopic
observations with the SMARTS 1.5-m telescope at Cerro Tololo Interamerican
Observatory and its Ritchey-Chretien spectrograph. Spectra were obtained by the
SMARTS service observers on 2012 May~28 (UT) with grating setup 26/Ia (covering
3660--5440~\AA\ at a resolution of 4.3~\AA), and on 2012 September~9 with
grating setup 47/Ib (5650--6970~\AA, resolution 3.1~\AA)\null. Exposure times
were $3\times300$~s and $3\times400$~s, respectively. A spectrophotometric
standard star was also observed on both nights (LTT~4364 and Feige~110,
respectively), allowing flux calibration of the NSV~11749 spectra. Since the
second night was reported to be non-photometric, we have scaled the spectrum
from that night to match the extrapolated flux level of the earlier blue
spectrum.

The optical spectra are plotted in Figure~2. The top panel shows the spectrum
scaled to show the emission lines, and the bottom panel expands the same data to
show the continuum. If NSV~11749 had been a born-again event like V605~Aql or
V4334~Sgr, we would expect to see an extremely hydrogen-deficient object; for
example, at the present time V605~Aql shows strong emission lines of
[\ion{O}{3}], a complete absence of Balmer emission, and weak emission features
of \ion{C}{4} arising from a dust-obscured hot Wolf-Rayet planetary-nebula
nucleus (e.g., Clayton et al.\ 2006).

Instead, NSV~11749 shows very strong emission lines of the Balmer series, along
with \ion{He}{1} and \ion{He}{2}\null. There are no forbidden lines detected.
The spectrum also shows a broad emission feature at about 6825~\AA\null.
Emission at $\lambda$6825 is a feature seen only in symbiotic binaries (e.g.,
Webster \& Allen 1975; Allen 1980), and has been identified with emission from
the strong \ion{O}{6} 1032~\AA\ resonance line that has been Raman-scattered by
neutral hydrogen (Schmid 1989, 2001; Lee \& Kang 2007). There is thus little
doubt that NSV~11749 is a symbiotic star, in which a hot compact star, usually a
WD, accretes from the wind of a late-type red giant. The overall pattern of
emission lines in NSV~11749 is fairly similar to that seen (in quiescence) in
the symbiotic nova V1329~Cyg (Kenyon \& Fernandez-Castro 1987, Fig.~1b; Munari
\& Zwitter 2002, Figs.~192 and 193).

The bottom panel in Figure~2 zooms in on the continuum, which, although noisy,
clearly shows molecular bands of TiO, indicative of an early M spectral type. To
confirm this classification, we superpose a scaled spectrum of the M1~III
standard star $\sigma$~Virginis (HD~115521), obtained from the Munari \& Zwitter
(2002) atlas\footnote{The data were downloaded from
\url{http://ulisse.pd.astro.it\slash symbio\_atlas}}, which provides a
reasonable match. We estimate a reddening of NSV~11749 of $E(B-V)\simeq0.75$
(based on Schlegel, Finkbeiner, \& Davis 1998 as well as the discussion in the
next paragraph), and we applied this amount to the $\sigma$~Vir spectrum, using
the formulae of Cardelli, Clayton, \& Mathis (1989).

On 2012 September 20, we obtained a near-infrared (NIR) spectrum of NSV~11749
with the Folded-port InfraRed Echellette spectrograph (FIRE; Simcoe et al.\
2008, 2010) on the 6.5-m Magellan Baade Telescope. We used the low-dispersion,
high-throughput prism mode, and completed an ABBA dither sequence.  Each
integration was in ``Sample Up the Ramp'' mode (10.6~s). The spectra span
0.8--2.5~$\mu$m at a resolution ranging from 300--500. Immediately afterwards,
we obtained a spectrum of the nearby A0~V star HD~177880B (HIP~93835) for the
purposes of flux calibration and removal of telluric absorption features, as
described by Vacca, Cushing, \& Rayner (2003). Data were reduced using the
FIREHOSE pipeline developed by R.~Simcoe, J.~Bochanski, and M.~Matejek. The
resulting spectrum is shown as a black line in Figure~3.

The NIR spectrum of NSV~11749 shows emission lines of the Paschen series and
\ion{He}{1} 10830~\AA, superposed on a late-type absorption spectrum.  We
compared the absorption-line spectrum to a library\footnote{Available at
\url{http://irtfweb.ifa.hawaii.edu/$^\sim$spex/IRTF\_Spectral\_Library/}} of
standard stellar spectra obtained with the IRTF (Cushing, Rayner, \& Vacca 2005;
Rayner, Cushing, \& Vacca 2009). We find a reasonable match of NSV~11749 with
the M2~III standard star 87~Virginis (HD~120052), with an extinction of
$E(B-V)=0.75$ applied to 87~Vir, as plotted in red in Figure~3. There is an
apparent broad emission line at 1.999~$\mu$m, which we have been unable to
identify; it may be an artifact due to incomplete telluric feature removal.

% 10/14/12: checking whether the 1.999 um line could be Raman-scattered Ne V,
% which seems to fit at first glance. Young et al. (2005, ApJ 618, 891, Table 2)
% list emission lines in the FUSE spectrum of AG Dra:
% 
% 		2000 Mar	2001 Apr
% 
% O VI 1031	5290		3930
% O VI 1037	1750		1340
% Ne V 1136	  27	     	  18
% Ne V 1145	  62		  45  <= they derived wl of 1145.615
% 
% This gives a predicted Raman w.l. of 1/((1/1145.615)-(1/1215.67))
%          = 19879.9 A.  
% The 6825 line is at 1/((1/1031.928)-(1/1215.67)) = 6827.421 A (vac) 
% 						 = 6825.534 (air).
% Observed feature is broad, from about 6818.2 to 6836.9, avg=6827.55
% 				vel = -322       +499        +88
% 
% so the 1.99 um prediction is:  1.98585 to 1.991298  center 1.98857 
%                        obs'd   1.992      2.045       1.998
% so it's pretty far off the predicted wl, and observed line is much wider than
% expected....

% ....10/15/12: but an email from Mansi with her arc line list shows that she is
% in fact using air wavelengths.  So the air w.l. of 1.998 corresponds to vacuum
% w.l. = 19985.4. But this doesn't change the conclusion that the match isn't
% good.

\section{Conclusion: Symbiotic Nova, Not Born-Again Red Giant}

Our observations verify that NSV~11749 is a symbiotic binary, containing a
compact hot object and a companion red giant with a spectral type of about
M1--2~III\null. Its location in the IPHAS $r'-H_\alpha$ vs.\ $r'-i'$ diagram
(Corradi et al.\ 2008) is consistent with a ``stellar (S)''-type symbiotic,
rather than a ``dusty (D)'' type, as is its lack of a strong mid-IR excess in
the {\it WISE\/} photometry.

The eruption of NSV~11749 around 1903 may have been a classical symbiotic
outburst, but both its large amplitude and long duration are more suggestive of
a symbiotic nova (due to a thermonuclear runaway on the WD component). Less than
a dozen symbiotic novae have been observed to date (e.g., Miko{\l}ajewska 2010;
Tang et al.\ 2012). It would be of interest to determine the orbital period and
other properties of this apparent new member of this rare class.

\acknowledgments

This work was partially supported by the STScI Director's Discretionary Research
Fund. We thank Marcelo Miller Bertolami, Ulisse Munari, and David Williams for
useful comments, Fred Walter for scheduling the SMARTS 1.5m observations, and
Manuel Hern\'andez and Rodrigo Hern\'andez for making the observations. MMK
acknowledges generous support from the Hubble Fellowship and Carnegie-Princeton
Fellowship programs. This research has made use of the SIMBAD database, operated
at CDS, Strasbourg, France. This publication makes use of data products from the
{\it Wide-field Infrared Survey Explorer}, which is a joint project of UCLA, and
JPL/Caltech, funded by NASA.

{\it Facilities:} \facility{SMARTS 1.5m telescope, Magellan Baade Telescope}

\clearpage

\begin{figure}
\begin{center}
\includegraphics[width=3in]{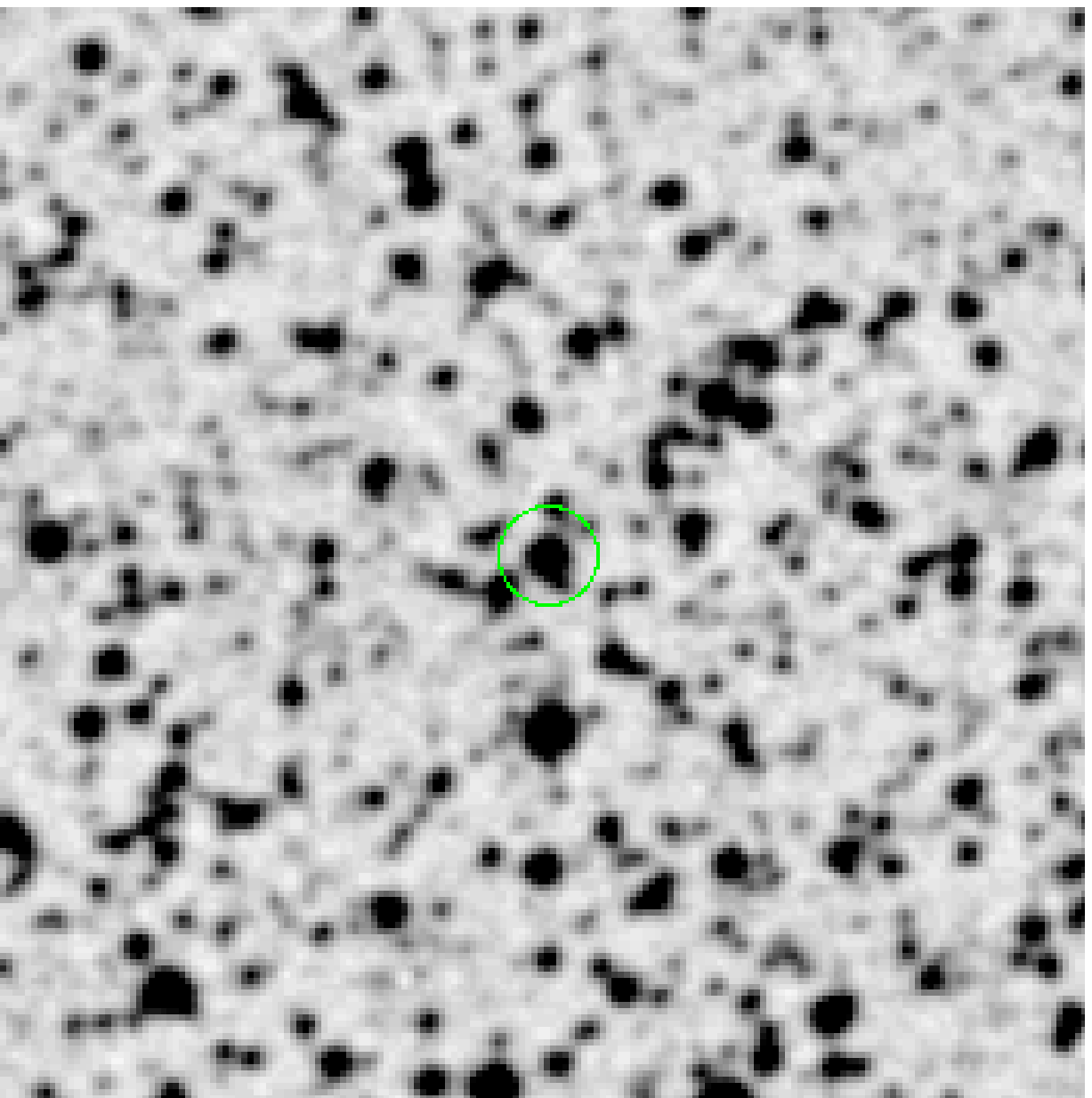}
\end{center} 
\figcaption{
Finding chart for NSV~11749, from the Digitized Sky Survey red image. The field
is $3'\times3'$ and north is at the top and east on the left. {\it The Digitized
Sky Surveys were produced at the Space Telescope Science Institute under U.S.
Government grant NAG W-2166.} 
}
\end{figure}

\begin{figure}
\begin{center}
\includegraphics[width=4.5in]{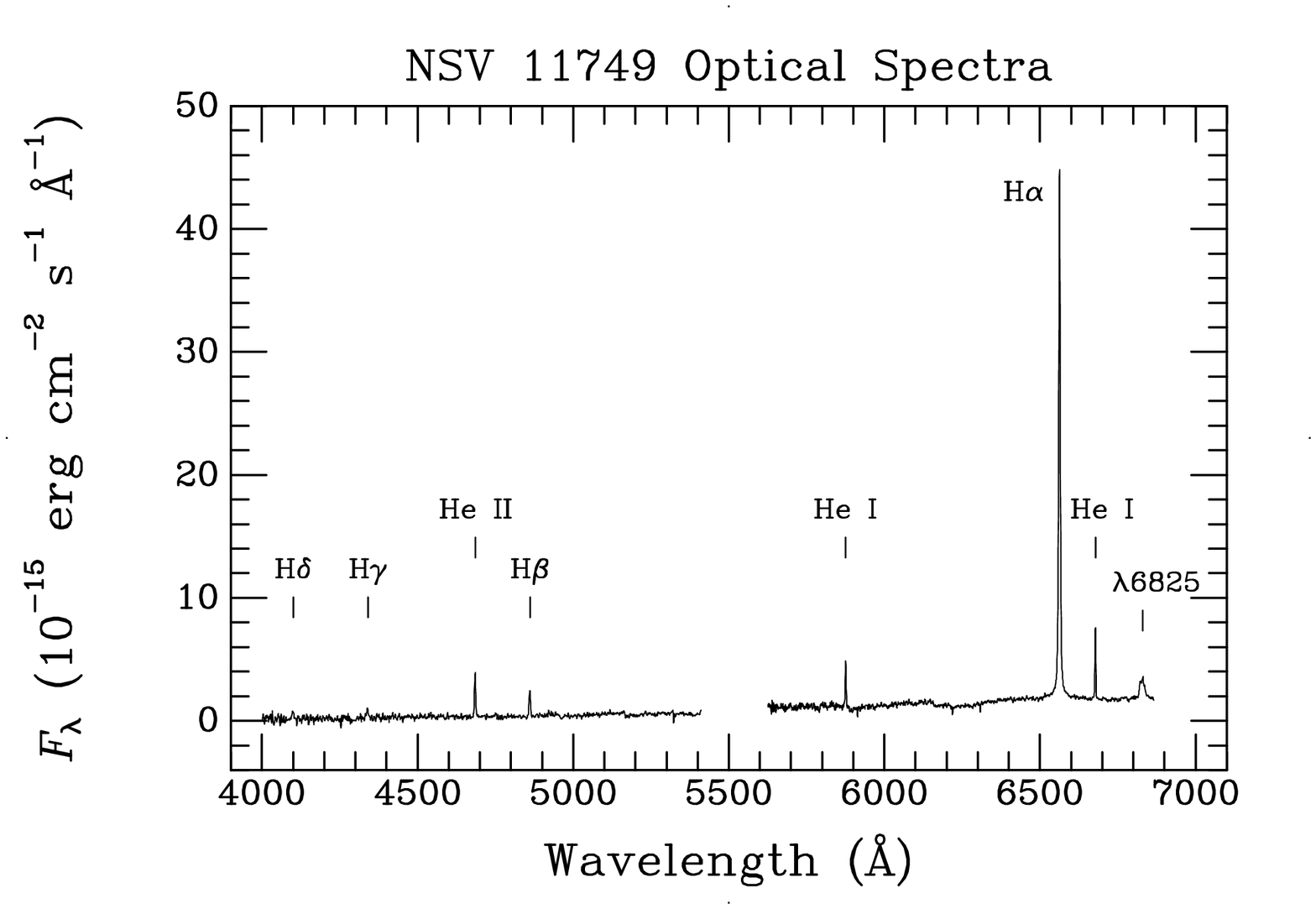}
\includegraphics[width=4.5in]{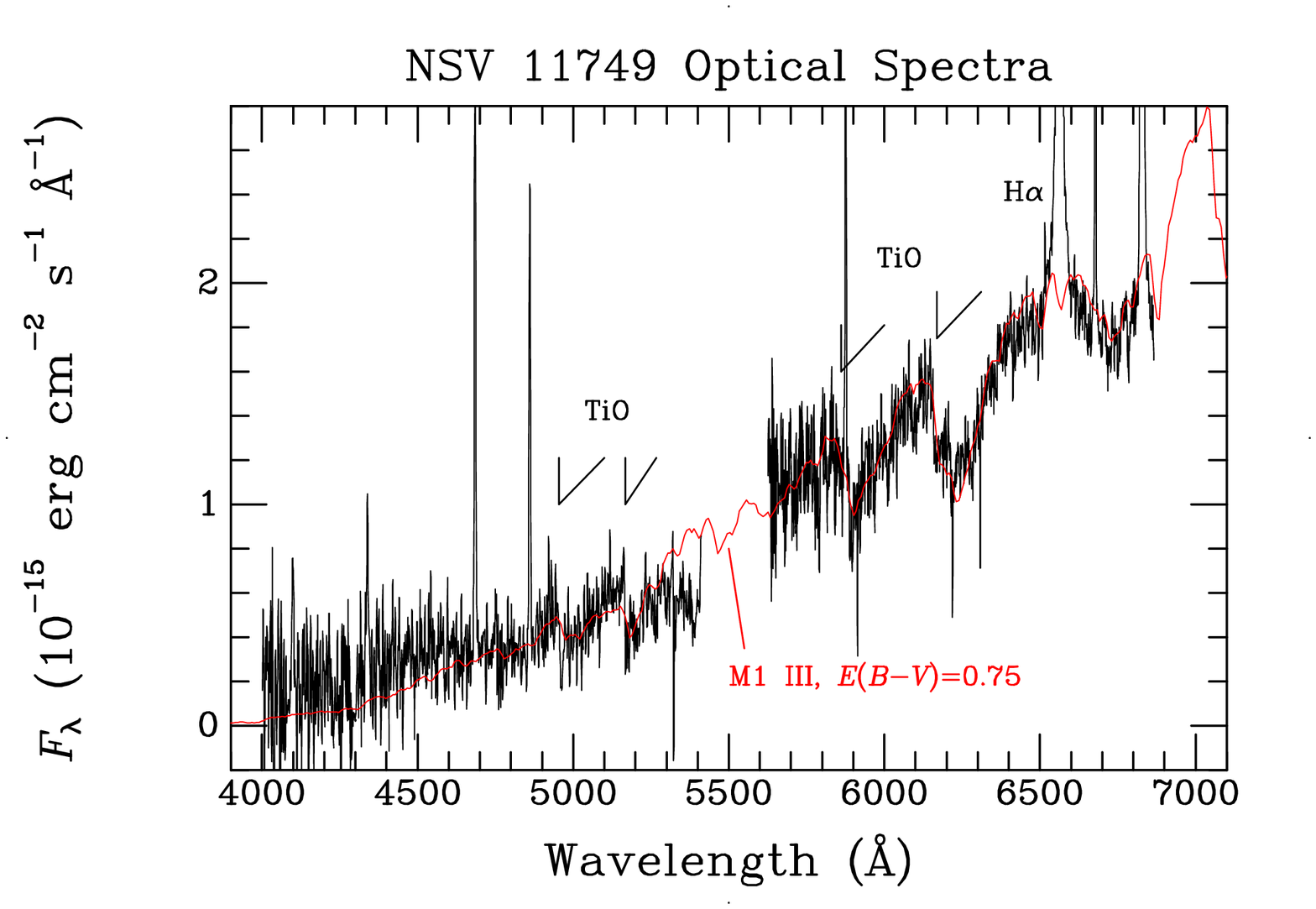}
\end{center} 
\figcaption{
SMARTS 1.5m optical spectra of NSV~11749 (black line), with a 3-point boxcar
smoothing applied. The spectra in the top panel are scaled to show the strong
emission lines of the Balmer series, He~I and II, and the broad $\lambda$6825
Raman feature characteristic of symbiotic binaries. Bottom panel scales the same
data to show TiO bands in the continuum, belonging to a cool companion of the
hot component of the binary. Superposed in red is a scaled spectrum of the
M1~III standard star $\sigma$~Vir, reddened by $E(B-V)=0.75$,  which provides a
reasonable match.
}
\end{figure}

\begin{figure}
\begin{center}
\includegraphics[width=6.5in]{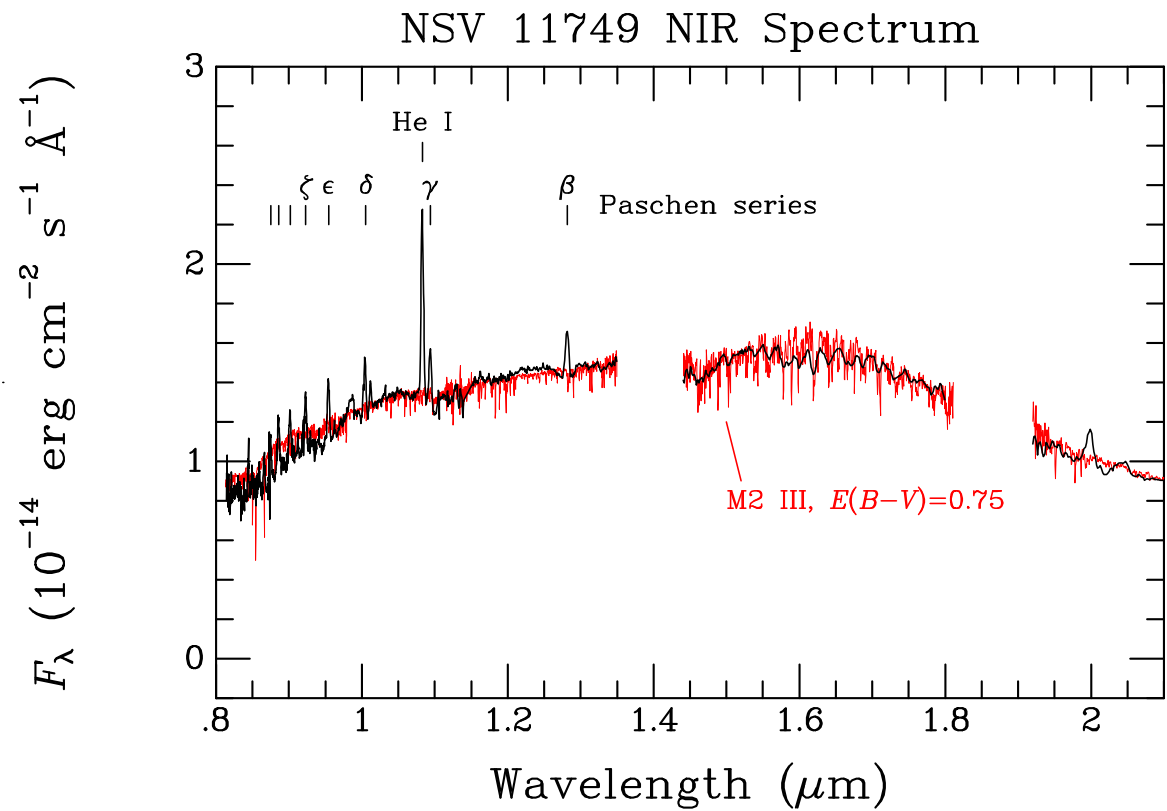}
\end{center} 
\figcaption{
Magellan NIR spectrum of NSV~11749 (black line), showing emission of the Paschen
series and \ion{He}{1} 1.0830~$\mu$m. Also plotted (red line) is the spectrum of
the M2~III standard star 87~Vir, reddened by $E(B-V)=0.75$ and scaled to the
flux of NSV~11749, which provides a good fit to the energy distribution.
}
\end{figure}

\end{document}